\documentclass{mem}
\usepackage{natbib}\usepackage{txfonts}\usepackage{balance}
\usepackage{graphicx}
\usepackage[a4paper,breaklinks,dvipdfm]{hyperref}
\idline{00}{00}
\begin{document}
\def\teff{$T\rm_{eff }$}
\def\kms{$\mathrm {km s}^{-1}$}

\title{
The strange case of NGC2419: stellar populations, chemical composition, dynamics.}

   \subtitle{}

\author{
M. \,Bellazzini\inst{1}, A. \,Mucciarelli\inst{2}, R.A. \,Ibata\inst{3}, A. \,Sollima\inst{4,1},
E. \,Dalessandro\inst{2}, S. \, Chapman\inst{5}, C. \,Nipoti\inst{2}, T. \,Merle\inst{6}, G. \,Beccari\inst{7}, C. \,Lardo\inst{2}, A. \,Bragaglia\inst{1}, E. \,Carretta\inst{1}, 
\and E. \,Pancino\inst{1}
          }

  \offprints{M. Bellazzini}

\institute{
Istituto Nazionale di Astrofisica --
Osservatorio Astronomico di Bologna, Via Ranzani 1,
I-40127 Bologna, Italy
\and
Dipartimento di Fisica e Astronomia, 
Universit\`a di Bologna, Viale Berti Pichat 6/2,
I-40127 Bologna, Italy
\and
Observatoire Astronomique, Universit\'e de Strasbourg, CNRS,
11 Rue de l'Universit\'e, F-6700 Strasbourg, France
\and
Istituto Nazionale di Astrofisica --
Osservatorio Astronomico di Padova, Vicolo dell'Osservatorio 5,
I-35122 Padova, Italy
\and
Institute of Astronomy, Madingley Road, Cambridge CB3 0HA, UK
\and
Universit\'e de Nice Sophia-antipolis, CNRS (UMR 7293), Observatoire de la C\^ote dÕAzur, Laboratoire Lagrange, BP 4229, 06304 Nice, France
\and
European Southern Observatory, Karl-Schwarzschild-Strasse 2, 85748 Garching bei M\"unchen, Germany
\\
\email{michele.bellazzini@oabo.inaf.it}
}

\authorrunning{Bellazzini et al.}

\titlerunning{The strange case of NGC2419}

\abstract{A few years ago we started an observational campaign aimed at the thorough study of the massive and remote globular cluster NGC2419. We have used the collected data, e.g., to test alternative theories of gravitation, to constrain the stellar M/L ratio by direct analysis of the observed luminosity function, and to search for Dark Matter within the cluster. Here we present some recent results about (a) the peculiar abundance pattern that we observed in a sample of cluster giants, and (b) newly found photometric evidence for the presence of multiple populations in the cluster. In particular, from new deep and accurate uVI LBT photometry, we find that the color spread on the Red Giant Branch is significantly larger than the observational errors both in V-I and u-V, and that the stars lying to the blue of the RGB ridge line are more concentrated toward the center of the cluster than those lying to the red of the ridge line.
\keywords{Stars: abundances -- Galaxy: globular clusters -- Globular clusters: individual: NGC~2419}
}
\maketitle{}

\section{Introduction}

The massive and remote globular cluster (GC) NGC2419 is a very interesting stellar system under several aspects \citep[see][and references therein]{mic12,dic_mult}. Five years ago, we started an observational campaign aimed at the thorough study of this cluster. We have used the collected data, e.g., to test alternative theories of gravitation \citep{iba11a,iba11b},
to constrain the stellar M/L ratio by direct analysis of the observed luminosity function \citep{mic12}, also providing a further observational confirmation of the un-relaxed dynamical status of the cluster \citep{ema}. 

Recently we have investigated the mass distribution within the cluster using both parametric and non-parametric methods, finding that the available data are not compatible with the presence of a ``cosmological'' Dark Matter halo surrounding the cluster \citep{iba12}.  

In the following we briefly report on the chemical composition of cluster stars and on newly found photometric evidences of the presence of multiple populations in NGC2419.

\section{Chemical composition}

In \citet{muccia} we have presented Fe, Ca, Ti, Mg, and K abundances for 49 Red Giant Branch (RGB) stars of NGC2419, derived from a subset of the high signal-to-noise medium-resolution spectra presented in \citet{iba11a}.

\begin{figure}[t!]
\resizebox{\hsize}{!}{\includegraphics[clip=true]{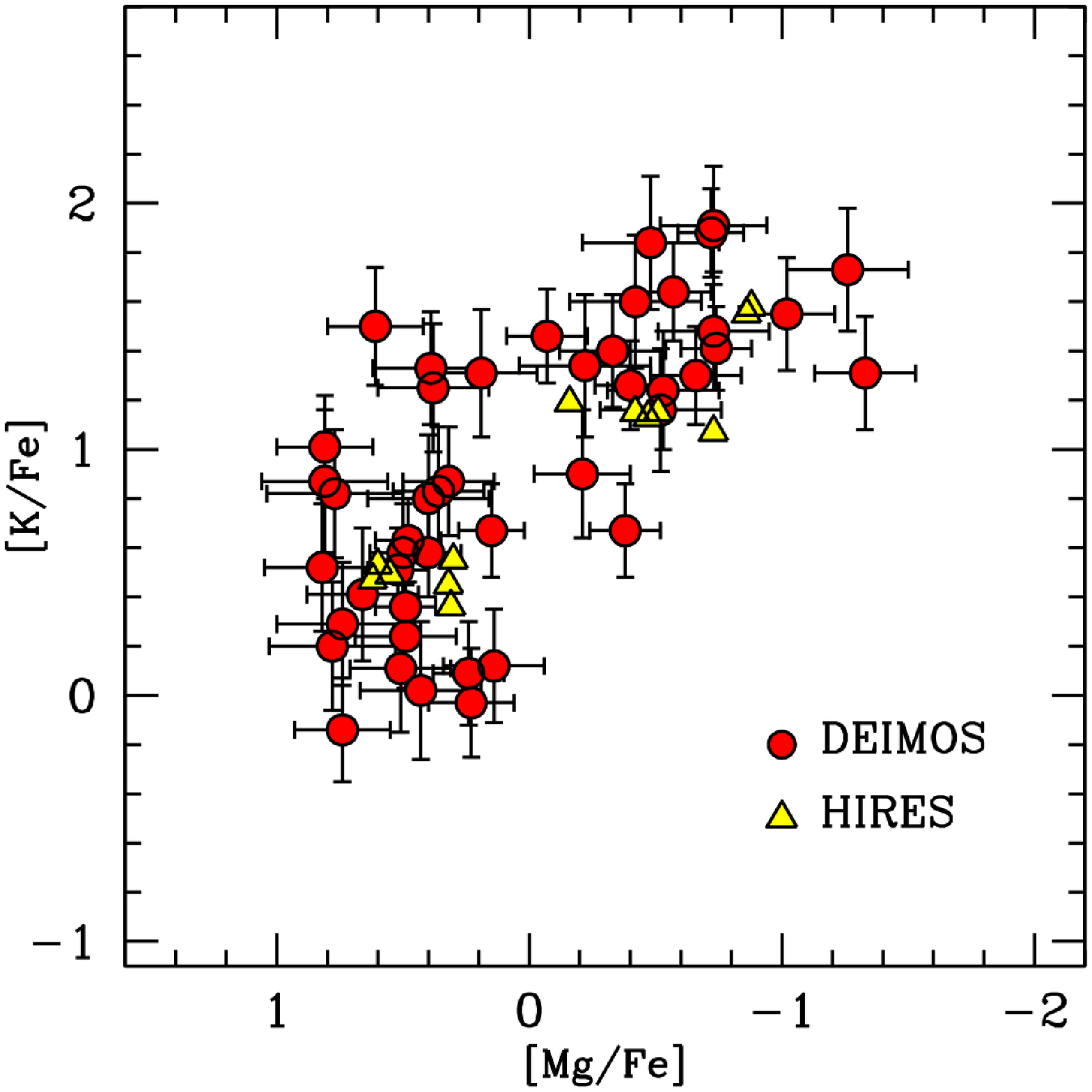}}
\caption{\footnotesize Potassium - Magnesium anti-correlation in NGC2419 Red Giants, from the analysis by 
\citet[red circles][]{muccia} and by \citet[][yellow triangles]{judy12}.
}
\label{mgk}
\end{figure}
\begin{figure}[t!]
\resizebox{\hsize}{!}{\includegraphics[clip=true]{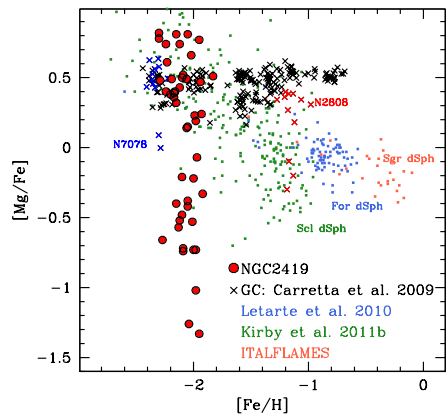}}
\caption{\footnotesize The distribution of [Mg/Fe] vs. [Fe/H] for NGC2419 stars \citep{muccia} is compared with (a) RGB stars for 15 Galactic GCs from \citet[][$\times$ symbols; in blue stars from NGC7078 and in red stars from NGC2808]{eugetut}, (b) stars from three dwarf spheroidal galaxies (small dots, see labels), from different sources \citep{k11a,k11b,leta,boni}.
}
\label{nanek}
\end{figure}

We found that cluster stars show no detectable spread in the abundance of Fe, Ti, and Ca, while they display a huge spread in both Mg and K abundances. In Fig.~\ref{mgk} we show that Mg and K abundances are anti-correlated and the distributions of both elements are clearly bimodal. 

Fig.~\ref{nanek}, on the other hand, show that (a) the spread in [Mg/Fe] detected in NGC2419 is larger than what found in any other GCs observed before, and (b) that the self-enrichment mechanism at work in the cluster is of completely different nature with respect to the one driving the chemical evolution of dwarf spheroidal galaxies. Our results have been nicely confirmed by \citet{judy12}, from high resolution spectra of 13 bright giants\footnote{The only difference in the results by \citet{muccia} and \citet{judy12} is about the spread in the Ca abundance: the former conclude that ``... the spread in calcium abundance is absent or very small...'', while \citet{judy12} claims that a small but real spread is actually there, especially among Mg-poor stars. From their data, using the Maximum Likelihood technique adopted by \citet{muccia}, we find that the intrinsic Ca spread is $\sigma_{[Ca/Fe]}=0.05 \pm 0.13$.}. 
We suggest that the observed abundance pattern may emerge as an extreme case of the multiple populations phenomenon in GCs \citep[see][for review and references]{grat12}. Within this framework, \citep{ventura} recently proposed an interesting theoretical scenario to interpret these results.

\section{Color spread in the RGB}

We acquired a large number of u$_{SDSS}$,V,I wide-field images of NGC2419, using simultaneously the two LBC cameras \citep{lbc} mounted at the binocular telescope LBT\footnote{\tt http://lbt.inaf.it/index.html}. We derived accurate V, u-V and V, V-I CMDs to search for color spreads on the RGB (especially in the NUV-optical colors), that are now generally recognized as a typical photometric signature of the presence of multiple populations \citep[see][]{lardo,grat12,sbor,milotuc}.
The original idea was to follow up the results by \citet{lardo}, that detected significant spread on the RGB of seven clusters, but were unable to draw firm conclusions on NGC2419 because the photometry they considered was not sufficiently accurate to study this very distant cluster.

\begin{figure}[t!]
\resizebox{\hsize}{!}{\includegraphics[clip=true]{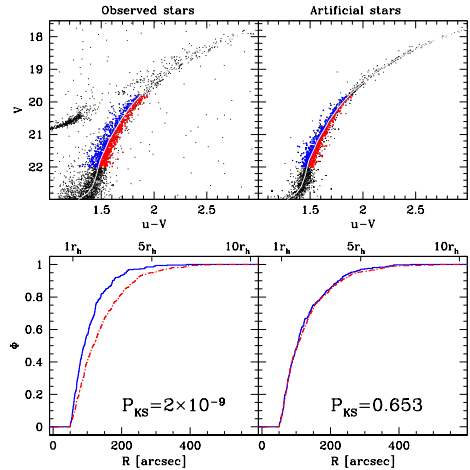}}
\caption{\footnotesize Upper panels: V, u-V CMDs of observed and artificial stars
from our LBT photometry of NGC2419. The RGB stars in the range $19.8\le V\le 22.0$
are divided in two samples, according to their position to the blue (RGB-B) or to the red (RGB-R) of the ridge line (in grey).
Lower panels: comparison between the radial distribution of RGB-B (continuous blue line) and 
RGB-R (dotted-dashed red line) for observed and artificial stars. The probability that RGB-R and RGB-B samples are drawn from the same radial distribution, according to a Kolmgorov-Smirnov test, is reported in the panels. Stars within $50$~arcsec from the cluster center have been excluded from the analysis. 
}
\label{radpost}
\end{figure}

We found that the observed color spread is significantly larger than what expected from observational errors (accurately traced by means of extensive artificial stars experiments), thus confirming that the effects of the presence of multiple populations can be detected on the RGB also in the case of NGC2419. While the spread is larger in u-V, it is clearly significant also in V-I, confirming earlier results by \citet{dic_mult}. These authors concluded that, given the low overall metallicity and the high degree of He enrichment occurred in this cluster , second generation stars lie to the blue of the RGB ridge line, at variance with the majority of other GCs \citep{lardo,grat12}. 

In particular, in Fig.~\ref{radpost} we demonstrate, for the first time, that second generation stars \citep[in the interpretation by][]{dic_mult} are more centrally concentrated than first generation stars.
The average fraction of second generation stars within the cluster, as derived from RGB stars, is $\sim 40$\%. A thorough analysis of this new photometric dataset will be presented in \citet{jack}.

\begin{acknowledgements}
We acknowledge the financial support of INAF through the PRIN-INAF 2009 grant assigned to the project {\em Formation and evolution of massive star clusters}, P.I. R.G. Gratton.
\end{acknowledgements}

\bibliographystyle{aa}

\end{document}